\documentclass[prd,aps,showpacs,tightenlines]{revtex4}
\usepackage{graphicx}

\newcommand{\bear}{\begin{array}}  \newcommand{\eear}{\end{array}}
\newcommand{\bea}{\begin{eqnarray}}  \newcommand{\eea}{\end{eqnarray}}
\newcommand{\beq}{\begin{equation}}  \newcommand{\eeq}{\end{equation}}
\newcommand{\bef}{\begin{figure}}  \newcommand{\eef}{\end{figure}}
\newcommand{\bec}{\begin{center}}  \newcommand{\eec}{\end{center}}

\newcommand{\bib}{\bibitem}

\def\IBIDD#1#2#3{{\it ibid}. {\bf #1}, #2 (20#3)}

\def\APJJ#1#2#3{Astrophys. J. {\bf #1}, #2 (20#3)}

\def\APJLL#1#2#3{Astrophys. J. Lett. {\bf #1}, L#2 (20#3)}

\def\JHEP#1#2#3{J. High Energy Phys. {\bf #1}, #2 (19#3)}

\def\NATT#1#2#3{Nature (London) {\bf #1}, #2 (20#3)}
\def\NPB#1#2#3{Nucl. Phys. {\bf B#1}, #2 (19#3)}
\def\NPBB#1#2#3{Nucl. Phys. {\bf B#1}, #2 (20#3)}

\def\PLB#1#2#3{Phys. Lett. B {\bf #1}, #2 (19#3)}
\def\PLBB#1#2#3{Phys. Lett. B {\bf #1}, #2 (20#3)}
\def\PLBold#1#2#3{Phys. Lett. {\bf#1B}, #2 (19#3)}

\def\PRD#1#2#3{Phys. Rev. D {\bf #1}, #2 (19#3)}
\def\PRDD#1#2#3{Phys. Rev. D {\bf #1}, #2 (20#3)}

\def\PRLL#1#2#3{Phys. Rev. Lett. {\bf#1}, #2 (20#3)}

\def\PRTT#1#2#3{Phys. Rep. {\bf#1}, #2 (20#3)}
\def\PTP#1#2#3{Prog. Theor. Phys. {\bf #1}, #2 (19#3)}

\def\RPP#1#2#3{Rep. Prog. Phys. {\bf #1}, #2 (19#3)}

\newcommand{\gtrsim}{ \mathop{}_{\textstyle \sim}^{\textstyle >} }
\newcommand{\lesssim}{ \mathop{}_{\textstyle \sim}^{\textstyle <} }
\newcommand{\ds}{\displaystyle}

\begin{document}

\title{Large lepton asymmetry from Q-balls}
\author{M. Kawasaki, Fuminobu Takahashi, and Masahide Yamaguchi}
\affiliation{Research Center for the Early Universe, University of Tokyo,
Tokyo 113-0033, Japan}
\date{\today}

\begin{abstract}
    We propose a scenario which can explain large lepton asymmetry and
    small baryon asymmetry simultaneously. Large lepton asymmetry is
    generated through the Affleck-Dine (AD) mechanism and almost all the
    produced lepton numbers are absorbed into Q-balls (L-balls). If the
    lifetime of the L-balls is longer than the onset of electroweak phase
    transition but shorter than the epoch of big bang nucleosynthesis
    (BBN), the large lepton asymmetry in the L-balls is protected from
    sphaleron effects. On the other hand, small (negative) lepton numbers
    are evaporated from the L-balls due to thermal effects, which are
    converted into the observed small baryon asymmetry by virtue of
    sphaleron effects. Large and positive lepton asymmetry of electron
    type is often requested from BBN. In our scenario by choosing an
    appropriate flat direction in the minimal supersymmetric standard
    model, we can produce positive lepton asymmetry of the electron type
    but totally negative lepton asymmetry.
\end{abstract}

\pacs{PACS numbers: 98.80.Cq}
\maketitle


\section{Introduction}

\label{sec:introduction}

The success of big bang nucleosynthesis (BBN) is one of the most powerful
pieces of evidence of standard big bang cosmology \cite{BBN}.  Roughly
speaking, the predicted primordial abundances of light elements (D,
$^3$He, $^4$He, and $^7$Li) coincide with those inferred from observations
for the baryon-to-photon ratio $\eta \sim 5 \times 10^{-10}$. However, as
observations are improved and their errors are reduced, a small discrepancy
may appear \cite{BBN2}. Furthermore, $\eta$ is also determined by
observations of small scale anisotropies of the cosmic microwave
background radiation (CMB)\cite{BOOMERANG,MAXIMA,DASI,BM}, which may also
cause the small discrepancy. Of course, the discordance may be completely
removed as observations are further improved. However, it is also probable
that such small discrepancies are genuine and suggest additional physics
in BBN.

These discrepancies are often eliminated if predicted primordial abundance
of $^{4}$He is decreased.  Such a decrease is realized if there exists
large and positive lepton asymmetry of electron type \cite{KS}. This is
mainly because the excess of electron neutrinos shifts the chemical
equilibrium between protons and neutrons toward protons, which reduces the
predicted primordial abundance of $^{4}$He. Note that this effect is much
more effective than the corresponding speed up effect, that is, an
increase of the Hubble expansion due to the presence of the chemical
potential, which makes the predicted primordial abundance of $^{4}$He
increase. However, large and positive lepton asymmetry of the electron
type is incompatible with small baryon asymmetry if we take account of the
sphaleron effects, which convert lepton asymmetry to baryon asymmetry of
the same order with the opposite sign \cite{sphaleron}. This problem is
evaded if one of the following three conditions is satisfied, that is, (a)
lepton asymmetry is generated after electroweak phase transition but
before BBN, (b) sphaleron processes do not work, (c) positive lepton
asymmetry of the electron type is generated but no total lepton asymmetry
is generated.

The first condition was discussed in the context of neutrino oscillations
\cite{oscillation}. In this case, large lepton asymmetry is generated
through oscillations between active neutrinos and sterile neutrinos.  
In order to explore the second condition, one should note that
the presence of large chemical potential prevents restoration of
electroweak symmetry \cite{nonrestoration}. Based on this fact, large
lepton asymmetry compatible with small baryon asymmetry was discussed
\cite{second}.  The third condition is discussed by March-Russell {\it et
al.}  \cite{MRM}. The Affleck-Dine mechanism produces positive lepton
asymmetry of electron type but no total lepton asymmetry, that is, $L_{e}
= - L_{\mu} > 0$ and $L_{\tau} = 0$ for some flat direction, which
generates small baryon asymmetry due to thermal mass effects of sphaleron
processes.

In this paper we consider another possibility, which is something like the
combination of (a) and (b). The Affleck-Dine (AD) mechanism produces
positive lepton asymmetry of the electron type but totally negative lepton
asymmetry by choosing an appropriate flat direction in the minimal
supersymmetric standard model (MSSM) \cite{AD}.  As an example, we
identify the ``$e^{c}_{1}L_{2}L_{3}$'' flat direction to be the AD field
and consider the Affleck-Dine leptogenesis. Here subscripts represent the
generations. Then, $L_{e} = - L_{\mu} = - L_{\tau} = - L_{\rm total} > 0$
is realized. The shift of the chemical equilibrium between neutrons and
protons due to the positive chemical potential of electron neutrinos
affects the results of BBN dominantly, while the speed-up effect caused by
all the species of neutrinos is relatively negligible. After the
Affleck-Dine leptogenesis, the AD field experiences spatial instabilities
and deforms into nontopological solitons, Q-balls
(L-balls)~\cite{Kusenko,Enqvist,Kasuya1}. Then, almost all the produced
lepton numbers are absorbed into the L-balls \cite{Kasuya1,Kasuya3}. If
the lifetime of such L-balls is longer than the onset of electroweak phase
transition but shorter than the epoch of BBN, the large lepton asymmetry
is protected from sphaleron effects and later released into the universe
by the decay of the L-balls. On the other hand, small (negative) lepton
numbers are evaporated from the L-balls due to thermal effects before the
electroweak phase transition, which are transformed into small baryon
asymmetry through the sphaleron effect.

In our scenario we consider the Affleck-Dine mechanism and the
subsequent Q-ball formation in the gauge-mediated supersymmetry (SUSY)
breaking model. This is mainly because, in the gravity-mediated SUSY
breaking model, the energy per unit charge of Q-balls is large enough
to produce the lightest supersymmetric particles (LSPs) so that they
will overclose the universe.
Therefore, we do not consider the gravity-mediated SUSY breaking model.
Since we assume that the AD field starts oscillating from the
gravitational scale to produce large lepton asymmetry, the produced
Q-balls are ``new'' \cite{Kasuya2} or ``delayed''- type \cite{Kasuya3},
depending on the sign of the coefficient of the one-loop correction to the
effective potential. However, since the decay processes of the new type
Q-balls are not completed before BBN, our scenario does not apply for
them.  Thus, we concentrate on delayed-type Q-balls in gauge-mediated SUSY
breaking models.

The rest of the paper is as follows. In Sec. \ref{sec:ADmechanism}, we
briefly review the Affleck-Dine mechanism and properties of Q-balls. In
Sec.  \ref{sec:lepton}, we discuss our mechanism to generate large lepton
asymmetry compatible with small baryon asymmetry. Section \ref{sec:con} is
devoted to discussion and conclusions.

\section{Affleck-Dine mechanism and Q-ball formation}

\label{sec:ADmechanism}

In this section we briefly review the Affleck-Dine mechanism and
properties of Q-balls. In MSSM, there exist flat directions, along which
there are no classical potentials in the supersymmetric limit. Since flat
directions consist of squarks and/or sleptons, they carry baryon and/or
lepton numbers, and can be identified as the Affleck-Dine (AD) field. In
the following discussion, we adopt the ``$e^{c}LL$'' direction as the AD
field. In this case the AD field carries only the lepton number.

These flat directions are lifted by supersymmetry (SUSY) breaking
effects. In the gauge-mediated SUSY breaking model, the potential of a
flat direction is parabolic at the origin, and almost flat beyond the
messenger scale \cite{Kusenko,Kasuya3,Gouvea},
\begin{equation}
    V_{gauge} \sim \left\{ 
      \begin{array}{ll}
          m_{\phi}^2|\Phi|^2 & \quad (|\Phi| \ll M_S), \\
          \ds{M_F^4 \left(\log \frac{|\Phi|^2}{M_S^2} \right)^2}
          & \quad (|\Phi| \gg M_S), \\
      \end{array} \right.
\end{equation}
where $m_{\phi}$ is a soft breaking mass $\sim$ O(1 TeV), $M_F$ is the
SUSY breaking scale, and $M_{S}$ is the messenger mass scale.

Since gravity always exists, flat directions are also lifted by
gravity-mediated SUSY breaking effects \cite{EnqvistMcDonald98},
\begin{equation}
    V_{grav} \simeq m_{3/2}^2 \left[ 1+K
      \log \left(\frac{|\Phi|^2}{M^2} \right)\right] |\Phi|^2,
\end{equation}
where $K$ is the numerical coefficient of the one-loop corrections and $M$
is the gravitational scale ($\simeq 2.4 \times 10^{18}$ GeV). This term can
be dominant only at high energy scales because of small gravitino mass
$\lesssim O(1\mbox{ GeV})$.

There is also the thermal effect on the potential, which appears at
two-loop order as pointed out in Ref. \cite{AnisimovDine}. This effect
comes from the fact that the running of the gauge coupling $g(T)$ is
modified by integrating out heavy particles which directly couple with
the AD field. This contribution to the effective potential is given by
\begin{equation}
    V_T^{(2)} \sim c~ \alpha_{\rm w}^2 T^4 \log\frac{|\Phi|^2}{T^2},
\end{equation}
where $|c| \sim 1$, and $\alpha_{\rm w} \equiv g_{\rm w}^2/4 \pi$
represents the gauge coupling constant of the weak interaction since we
consider $e^c LL$ direction. Though the sign of $c$ depends on flat
directions, it is irrelevant to our discussion since we assume that the
zero-temperature potential dominates over the thermal effects. Note that
$\alpha_{\rm w}$ should be replaced with $\alpha_{s}$ for those flat
directions which contain squarks.

The lepton number is usually created just after the AD field starts
coherent rotation in the potential, and its number density $n_L$ is
estimated as
\begin{equation}
    n_L(t_{osc}) \simeq \varepsilon \omega \phi_{osc}^2,
\end{equation}
where $\varepsilon(\lesssim 1)$ is the ellipticity parameter, which
represents the strongness of the A term, and $\omega$ and $\phi_{osc}$
are the angular velocity and amplitude of the AD field at the
beginning of the oscillation (rotation) in its effective potential.

Actually, however, the AD field experiences spatial instabilities during
its coherent oscillation, and deforms into nontopological solitons called
Q-balls \cite{Kusenko,Enqvist,Kasuya1}. When the zero-temperature
potential $V_{gauge}$ dominates at the onset of coherent oscillation of
the AD field, the gauge-mediation type Q-balls are formed. Their mass $M_{Q}$
and size $R_Q$ are given by \cite{Dvali}
\begin{equation}
    \label{eq:mass}
    M_Q \sim M_F Q^{3/4}, \qquad R_Q \sim M_F^{-1} Q^{1/4}.
\end{equation}
From the numerical simulations \cite{Kasuya1,Kasuya3}, the produced
Q-balls absorb almost all the charges carried by the AD field and the
typical charge is estimated as \cite{Kasuya3}
\begin{equation}
    Q \simeq \beta \left(\frac{\phi_{osc}}{M_F}\right)^4
\end{equation}
with $\beta \approx 6 \times 10^{-4}$.

There are also other cases where $V_{grav}$ dominates the potential at
the onset of coherent oscillation of the AD field. If the coefficient
of the one-loop correction $K$ is negative, the gravity-mediation type Q-balls
(``new'' type) are produced \cite{Kasuya2}. On the other hand, if $K$ is
positive, Q-balls do not form until the AD field leaves the $V_{grav}$
dominant region. Later it enters the $V_{gauge}$ dominant region and
experiences instabilities so that the gauge-mediation type Q-balls are
produced (delayed-type Q-balls) \cite{Kasuya3}. 

In our scenario described in the next section, the AD field starts to
oscillate from the gravitational scale, i.e., $\phi_{osc} = M$, which leads
to the formation of new or ``delayed''-type Q-balls. However, our scenario
does not work for new type Q-balls because the produced Q-balls are large
and do not decay before BBN. Hence, we concentrate on the delayed-type
Q-balls below. Since the sign of $K$ is in general indefinite and
dependent on the model of the messenger sector in gauge-mediated SUSY
breaking models, we assume that $K$ is positive and delayed-type Q-balls
are formed.

When the AD field starts to oscillate in the $V_{grav}$ dominant
region, where $H_{osc} \sim \omega \sim m_{3/2}$, the lepton number is
produced as $n_L \simeq \varepsilon m_{3/2}\phi_{osc}^2$. Since
the delayed-type Q-balls are formed only after the AD field enters the
$V_{gauge}$ dominant region for positive $K$, the charge of Q-ball is
given by
\begin{equation}
    \label{eq:delayq}
    Q \sim \beta \left(\frac{\phi_{eq}}{M_F}\right)^4
      \sim \beta \left(\frac{M_F}{m_{3/2}}\right)^4
\end{equation}
with $\phi_{eq} \sim M_F^2/m_{3/2}$. Here the subscript ``eq'' denotes a value
when the gauge- and the gravity-mediation potentials become equal. Thus
the delayed-type Q-balls are formed at $H_{eq} \sim M_F^2/M$.

As we mentioned above, Q-balls absorb almost all the charges carried by
the AD field. If we adopt the $e^{c}LL$ direction, all the lepton charges
are confined in the Q-balls, namely, L-balls.  Consequently, we must take
out lepton charge from the L-balls through the evaporation, diffusion, and
their decay. Part of the evaporated lepton charge is transformed into
baryon charge by the sphaleron process, which accounts for the present
baryon asymmetry.

In the case of L-balls, they decay into leptons such as neutrinos via gaugino
exchanges. The decay rate of Q-balls is bounded as \cite{Coleman}
\begin{equation}
    \label{eq:qdecay}
    \left|\frac{dQ}{dt}\right| \lesssim \frac{\omega^{3} A}{192 \pi^{2}},
\end{equation}
where $A$ is a surface area of the Q-ball. For L-balls, the decay rate is
estimated as a value of the order of the upper limit. 

According to Refs.\cite{Kasuya3,Laine,Banerjee}, we evaluate the
evaporation rate of L-balls, which is given by \cite{Laine}
\begin{eqnarray}
\label{eq:evap}
\zeta_{\rm evap} \equiv \frac{dQ}{dt}&=&-\kappa (\mu_{Q}-\mu_{\rm
plasma}) T^2 4 \pi R_{Q}^{2},\nonumber\\
&\simeq& - 4 \pi \kappa \mu_{Q} T^2 R_{Q}^{2} 
~~~{\rm for~~}\mu_{Q} \gg \mu_{\rm plasma} , 
\end{eqnarray}
where $\mu_{Q}$ and $\mu_{\rm plasma}$ are chemical potentials of the
Q-ball and plasma, and the coefficient $\kappa \lesssim 1$ includes
statistical and other numerical factors. The chemical potential of the
Q-ball is given as $\mu_{Q} \simeq \omega$ since the energy of the $\phi$
particle inside the Q-ball is $\omega$.  At $T \gtrsim m_{\phi}$, large
numbers of the scalar particles building up Q-balls are in the plasma,
which implies $\kappa \sim 1$. On the other hand, at $T \lesssim
m_{\phi}$, the evaporation from  Q-balls is suppressed by the
Boltzmann factor. In the case of L-balls, the main process of the
evaporation is $\phi \phi \rightarrow ll$ through wino or bino exchange,
which yields $\kappa \sim \alpha_{\rm w}^{2} T^2 /m_{\phi}^2$ at $T \lesssim
m_{\phi}$. 

However, if the charge transport is not effective enough, the
evaporated lepton charges in the ``atmosphere'' of the L-ball will
establish chemical equilibrium there. In this case, the dissipation of
the charge is determined by the diffusion. The diffusion rate is
estimated as \cite{Banerjee},
\begin{eqnarray}  
 \label{eq:diff}
\zeta_{\rm diff} \equiv \frac{dQ}{dt} &=& -4 \pi D R_{Q} \mu_{Q} T^2
\nonumber\\
         &\simeq& -4 \pi D T^2,
\end{eqnarray}
where the diffusion constant $D$ of relativistic sleptons and leptons in a
hot plasma is given by $D \simeq a/T$ with $a \sim 20$
\cite{Weldon,Nelson}. In short, the time scale of the charge
transportation is determined by the evaporation rate when $|\zeta_{\rm
evap}| < |\zeta_{\rm diff}|$, and by the diffusion rate when
$|\zeta_{\rm evap}| > |\zeta_{\rm diff}|$.

The amount of the evaporated charges can be estimated by integrating
Eqs.~(\ref{eq:evap}) and (\ref{eq:diff}) in the course of the evolution of
the universe. When the AD field starts to oscillate at the gravitational
scale, its oscillation energy is comparable to the total energy of the
universe. Therefore, the energy of the universe will be dominated by the
AD condensate or Q-balls soon after the reheating and the universe
continues to be matter dominated. The thermal history of the universe is
rather involved because radiation comes from both decays of an inflaton
and Q-balls. However, in fact, we have only to consider two cases where
the cosmic temperature decreases monotonically.

\section{Large Lepton Asymmetry from L-ball}

\label{sec:lepton}

In this section we give a detailed explanation of our scenario. Our goal
is to generate small baryon asymmetry and positive large lepton asymmetry
of the electron type simultaneously. In general, however, this is
difficult to accomplish because the chemical equilibrium induced by the
sphaleron transition forces the baryon and the lepton asymmetries to be
of the same order with opposite sign \cite{sphaleron}.  Hence we must get
over two problems : (i) how to protect large lepton asymmetry from being
converted to baryon asymmetry by the sphaleron process, and (ii) how to
reconcile the opposite sign of baryon and lepton asymmetries.

We show that these two obstacles can be evaded by considering the
Affleck-Dine leptogenesis and the subsequent L-ball formation using
$e^{c}LL$ direction. First of all, we give the outline of our scenario
and the solution to the problem (i). Large lepton asymmetry can be
generated if the A terms, which make the AD field rotate in the effective
potential, originate from some K\"{a}hler potential with vanishing
superpotential. Then the AD field starts to oscillate with large initial
amplitude $\phi_{osc} \simeq M$ and ellipticity $\varepsilon \simeq 1$. As
spatial instabilities grow, delayed-type L-balls are formed and absorb
almost all charges carried by the AD field. It is essential to our
scenario that the lepton asymmetry confined in the L-balls is kept from
the sphaleron process. However, a small part of lepton charges confined
in the L-balls are evaporated due to thermal effects. Thus, lepton charges
evaporated until the electroweak phase transition ($T \gtrsim T_{C} \sim
300$ GeV) $\Delta Q_{ew}$ are partly converted to baryon asymmetry through
the sphaleron process, which explains the present small baryon
asymmetry. On the other hand, large lepton asymmetry comes out through the
decay of the remnant L-balls after the electroweak phase transition, which
must be completed before  BBN. Thus the small ratio $\Delta Q_{ew}/Q$
is the source of hierarchy between the baryon and the lepton asymmetries.

Next we give a solution to the problem (ii), that is, the sign of the
lepton asymmetry. What we want to generate is positive baryon asymmetry
and positive lepton asymmetry of electron type. However, the sphaleron
process converts positive lepton asymmetry into negative baryon
asymmetry. To surmount this problem, we adopt the $e^{c}_{1}L_{2}L_{3}$
direction as the AD field, which leads $L_{e} = - L_{\mu} = - L_{\tau} = -
L_{\rm total} > 0$. Thus the positive lepton asymmetry of the electron
type is generated, whilst total lepton asymmetry is necessarily negative
in order to have positive baryon asymmetry through the sphaleron
transition. At the epoch of BBN, charged leptons except electrons have
already disappeared through decay and annihilation processes.  Also,
because of the charge neutrality of the universe, lepton asymmetry stored
in electrons are comparable to baryon asymmetry, which is rather
small. Thus, there can exist large lepton asymmetry only in the neutrino
sector. For later use, we define the degeneracy parameter $\xi_{l}$ as the
ratio of the chemical potential to the neutrino temperature. The presence
of chemical potentials speed up the universe, which leads to an increase
in the $n/p$ ratio. However, its effect is negligible in comparison with
the effect of the shift of chemical equilibrium between protons and
neutrons due to the chemical potential of electron neutrino in the case of
$|\xi_{\nu_e}| = |\xi_{\nu_\mu}| = |\xi_{\nu_\tau}|$.

Now we give a quantitative estimate for our scenario. We assume that the
zero-temperature potential dominates, i.e., $V_{gauge} \gg V_{T}^{(2)}$,
at the formation of the delayed-type L-balls with $H \sim H_{eq}$ :
\begin{equation}
 \label{eq:potential}
\alpha_{\rm w}^2 T_{eq}^4 < M_F^4,
\end{equation}
where $T_{eq}$ is the temperature of the universe just before the
delayed-type L-balls are formed. As shown below, this constraint is
automatically satisfied for the cases we consider.

The delayed-type L-balls must decay before  BBN,
\begin{equation}
\tau_{Q}=\left(\frac{1}{Q} \left|\frac{dQ}{dt}\right|\right)^{-1}
\lesssim  1 {\rm sec},
\end{equation}
which leads to the constraint
\begin{equation}
\label{eq:bbndecay}
\frac{m_{3/2}}{10 {\rm MeV}} \gtrsim \left(\frac{M_{F}}{10 {\rm
TeV}}\right)^{4/5}\,.
\end{equation}
Here Eq.~(\ref{eq:qdecay}) is used. In order to estimate the baryon and
the lepton to entropy ratio, it is necessary to evaluate the entropy
production by the decay of the L-balls. The decay temperature of the
L-balls, $T_d$, is given by
\begin{eqnarray}
\label{eq:dectemp}
T_{d} &=& \left(\frac{90}{\pi^2 g_{*}}\right)^{1/4}
\sqrt{M \frac{M_{F} Q^{-5/4}}{48 \pi}}\nonumber \\
& \simeq & 1.3 {\rm MeV} \left(\frac{M_{F}}{10 {\rm TeV}}\right)^{-2}
\left(\frac{m_{3/2}}{10 {\rm MeV}}\right)^{5/2},
\end{eqnarray}
where $g_{*}=10.75$ counts the total number of effectively massless
degrees of freedom. 

Now we turn to an account of the total evaporated charge, $\Delta Q$, and
the evaporated charge at temperatures above the electroweak phase
transition, $\Delta Q_{ew}$. In fact we have only to consider the
following two cases. In the other cases, the temperature during the
presence of L-balls does not exceed $T_{C}$ so that the evaporated lepton
numbers are not converted into baryon numbers.

 

First we consider the case that the delayed-type L-balls are formed
before the reheating and decay after that (case A). This is realized
if the following two conditions are satisfied,
\begin{eqnarray}
\label{eq:case1}
M_F &>& T_{RH},\\
T_{RH} &>& T_d.
\end{eqnarray}
Then the temperature at the L-ball formation is given by $T_{eq}=\sqrt{M_F
T_{RH}}$, which automatically satisfies the requirement
(\ref{eq:potential}). The temperature of the universe is approximately
given as
\begin{equation}
\label{eq:temperature1}
T \simeq \left\{
\begin{array}{ll}
\ds{\left(T_{RH}^2 M H \right)^{1/4}} & {\rm for~} \ds{T_{RH} < T }, \\
\ds{\left(T_{RH}^{-1} M^2 H^2 \right)^{1/3}} & {\rm for~} \ds{T_p < T <
T_{RH}}, \\
\ds{\left(T_{d}^2 M H \right)^{1/4}} & {\rm for~} \ds{T_{d} < T <
T_{p}}, \\
\sqrt{H M}  & {\rm for~} \ds{T < T_{d}}, 
\end{array} 
\right.
\end{equation}
where $T_p \equiv \left(T_{RH} T_d^4 \right)^{1/5}$ denotes the
temperature when the radiation derived from the decay of the L-balls
dominate over those derived from the inflaton. Note that the cosmic
temperature decreases monotonically in this case.  The condition that the
chemical equilibrium induced by the sphaleron transition are well
established is given by
\begin{equation}
T_{eq}=\sqrt{M_F T_{RH}} > T_C.
\end{equation}
With the use of Eq. (\ref{eq:temperature1}), the evaporation rate with
respect to the temperature is estimated as
\begin{equation}
\left(\frac{dQ}{dT}\right)_{evap} \simeq \left\{
\begin{array}{ll}
\ds{4 \pi  \frac{T_{RH}^2 M}{M_F T^{3}}  Q^{1/4}} & 
{\rm for~} m_\phi,\,T_{RH}<T<T_{eq} ,\\
\ds{4 \pi \alpha_{\rm w}^2 \frac{T_{RH}^2 M}{m_\phi^2 M_F T }  Q^{1/4}} & 
{\rm for~}T_{RH}<T< m_\phi,\,T_{eq} ,\\
\ds{4 \pi  \frac{M}{ M_F \sqrt{T T_{RH}}}  Q^{1/4}} & 
{\rm for~} m_\phi,\,T_p<T<T_{RH} ,\\
\ds{4 \pi  \alpha_{\rm w}^2 \frac{M T^{3/2}}{m_\phi^2 M_F T_{RH}^{1/2}}  Q^{1/4}} & 
{\rm for~} T_p<T<m_\phi,\,T_{RH} ,\\
\ds{4 \pi  \frac{T_{d}^2 M}{M_F T^{3}}  Q^{1/4}} & 
{\rm for~} m_\phi,\,T_{d}<T<T_{p} ,\\
\ds{4 \pi \alpha_{\rm w}^2 \frac{T_{d}^2 M}{m_\phi^2 M_F T }  Q^{1/4}} & 
{\rm for~}T_{d}<T< m_\phi,\,T_{p}.
\end{array}
\right.
\label{eq:gmevap}
\end{equation}
On the other hand, the diffusion rate with respect to the temperature
is given by
\begin{equation}
\left(\frac{dQ}{dT}\right)_{diff} \simeq \left\{
\begin{array}{ll}
\ds{4 \pi a \frac{T_{RH}^2 M}{T^4}} & {\rm for~} T_{RH} < T < T_{eq},\\
\ds{4 \pi a \frac{M}{T_{RH}^{1/2}T^{3/2}}} & {\rm for~}T_p < T <
T_{RH},\\
\ds{4 \pi a \frac{T_{d}^2 M}{T^4}} & {\rm for~} T_{d} < T < T_{p}.
\end{array}
\right.
\label{eq:gmdiff}
\end{equation}
By integrating Eqs. (\ref{eq:gmevap}) and (\ref{eq:gmdiff}), the
evaporated charges $\Delta Q$ and $\Delta Q_{ew}$ are found to be
the same order, and given by
\begin{equation}
\label{eq:delqew}
\Delta Q \simeq \Delta Q_{ew}  \simeq \left\{
\begin{array}{ll}
\ds{\frac{4 \pi a}{3} \frac{T_{RH}^2 M}{ m_{\phi}^3} }& {\rm
for~} T_{RH} < m_{\phi} < T_{eq}, \\
\ds{8 \pi a \frac{M}{\sqrt{m_{\phi} T_{RH}}}} & {\rm for~}T_{p} <
m_{\phi} < T_{RH}, \\
\ds{\frac{4 \pi a}{3} \frac{T_{d}^2 M}{ m_{\phi}^3} }& {\rm
for~} T_{d} < m_{\phi} < T_{p},
\end{array} 
\right.
\end{equation}
where we have used $T_C \sim m_{\phi}$.

 

Next we consider the case where delayed-type L-balls are formed after
the reheating and the temperature decreases monotonically (case B).
This is realized if the following conditions are satisfied,
\begin{eqnarray}
\label{eq:case2}
T_{RH} &>& M_{F},\\
M_{F} &>& \left(T_{RH}^2 T_d^3 \right)^{\frac{1}{5}},
\end{eqnarray}
Then the temperature at the Q-ball formation is given by
$T_{eq}=\left(M_F^4/T_{RH} \right)^{1/3}$, which again satisfies the
requirement (\ref{eq:potential}). Though the time evolution of the cosmic
temperature is the same as case A, the requirement for the sphaleron
process to work now reads
\begin{equation}
T_{eq}=\left(M_F^4/T_{RH} \right)^{1/3} > T_C.
\end{equation}
The evaporated charges $\Delta Q$ and $\Delta Q_{ew}$ can be estimated
similarly and given by
\begin{equation}
\label{eq:delqew2}
\Delta Q \simeq \Delta Q_{ew}  \simeq \left\{
\begin{array}{ll}
\ds{8 \pi a \frac{M}{\sqrt{m_{\phi} T_{RH}}}} & {\rm for~}T_{p} <
m_{\phi} < T_{eq}, \\
\ds{\frac{4 \pi a}{3} \frac{T_{d}^2 M}{ m_{\phi}^3} }& {\rm
for~} T_{d} < m_{\phi} < T_{p}
\end{array} 
\right..
\end{equation}

Finally we estimate the baryon (lepton) to entropy ratio, using
the results derived above. The baryon to entropy ratio is then given
by
\begin{eqnarray}
\label{eq:nbs}
&&\frac{n_{B}}{s} = \frac{8}{23} \frac{m_{3/2} M^2 }{
                    \frac{2 \pi^2}{45} g_{*} T_{d}^3} 
                    \frac{\frac{\pi^2}{90}g_* T_d^4}{ m_{3/2}^2} 
                    \frac{\Delta Q_{ew}}{Q},\nonumber \\ 
               &&~~~~     =\frac{2}{23} \frac{T_d}{m_{3/2}} 
                    \frac{\Delta Q_{ew}}{Q},\nonumber \\
               &\sim& \left\{
\begin{array}{ll}
\ds{4 \times 10^{-11} \left(\frac{m_{\phi}}{1{\rm
TeV}}\right)^{-3} \left(\frac{m_{3/2}}{1{\rm GeV}}\right)^{\frac{11}{2}}
\left(\frac{T_{RH}}{10{\rm GeV}}\right)^{2}
\left(\frac{M_F}{10^6{\rm GeV}}\right)^{-6}} & \\
~~~~~~~~~~~~~~~~~~~~~~~~\qquad\qquad\qquad\qquad\qquad\qquad\qquad
 {\rm for~} \ds{T_{RH} < m_{\phi} <  T_{eq}} & 
({\rm case A})\\
\ds{
3 \times 10^{-11} \left(\frac{m_{\phi}}{1{\rm
TeV}}\right)^{-\frac{1}{2}} \left(\frac{m_{3/2}}{1{\rm GeV}}\right)^{\frac{11}{2}}
\left(\frac{T_{RH}}{10^7{\rm GeV}}\right)^{-\frac{1}{2}}
\left(\frac{M_F}{3 \times 10^6{\rm GeV}}\right)^{-6}
} & \\ 
 ~~~~~~~~~~~~~~~~~~~~~~~~\qquad\qquad\qquad\qquad\qquad\qquad\qquad
{\rm for~} \ds{T_{p} < m_{\phi} <T_{eq}}& ({\rm case B})
\end{array} 
\right. ,\nonumber \\
&&
\end{eqnarray}
where we have used Eqs. (\ref{eq:delayq}), (\ref{eq:dectemp}),
(\ref{eq:delqew}), and (\ref{eq:delqew2}). Also we have assumed the
maximal CP violation. In the same way, the lepton number to entropy
ratio is given by
\begin{eqnarray}
\label{eq:nls}
\frac{n_{L}}{s} &=&- \frac{T_d}{4 m_{3/2}} \nonumber \\
               &\sim& -0.01 \times   \left(\frac{m_{3/2}}{1{\rm
GeV}}\right)^{\frac{3}{2}} \left(\frac{M_F}{5 \times 10^5 {\rm GeV}}\right)^{-2},
\end{eqnarray}
which yields \cite{KS}
\begin{eqnarray}
\xi_{\nu_e} &\simeq& - 10 \times \frac{n_{L}}{s}\nonumber \\
&\sim& 0.1 \times   \left(\frac{m_{3/2}}{1{\rm
GeV}}\right)^{\frac{3}{2}} \left(\frac{M_F}{5 \times 10^5{\rm GeV}}\right)^{-2}.
\end{eqnarray}

\begin{figure}
\includegraphics[width=10cm]{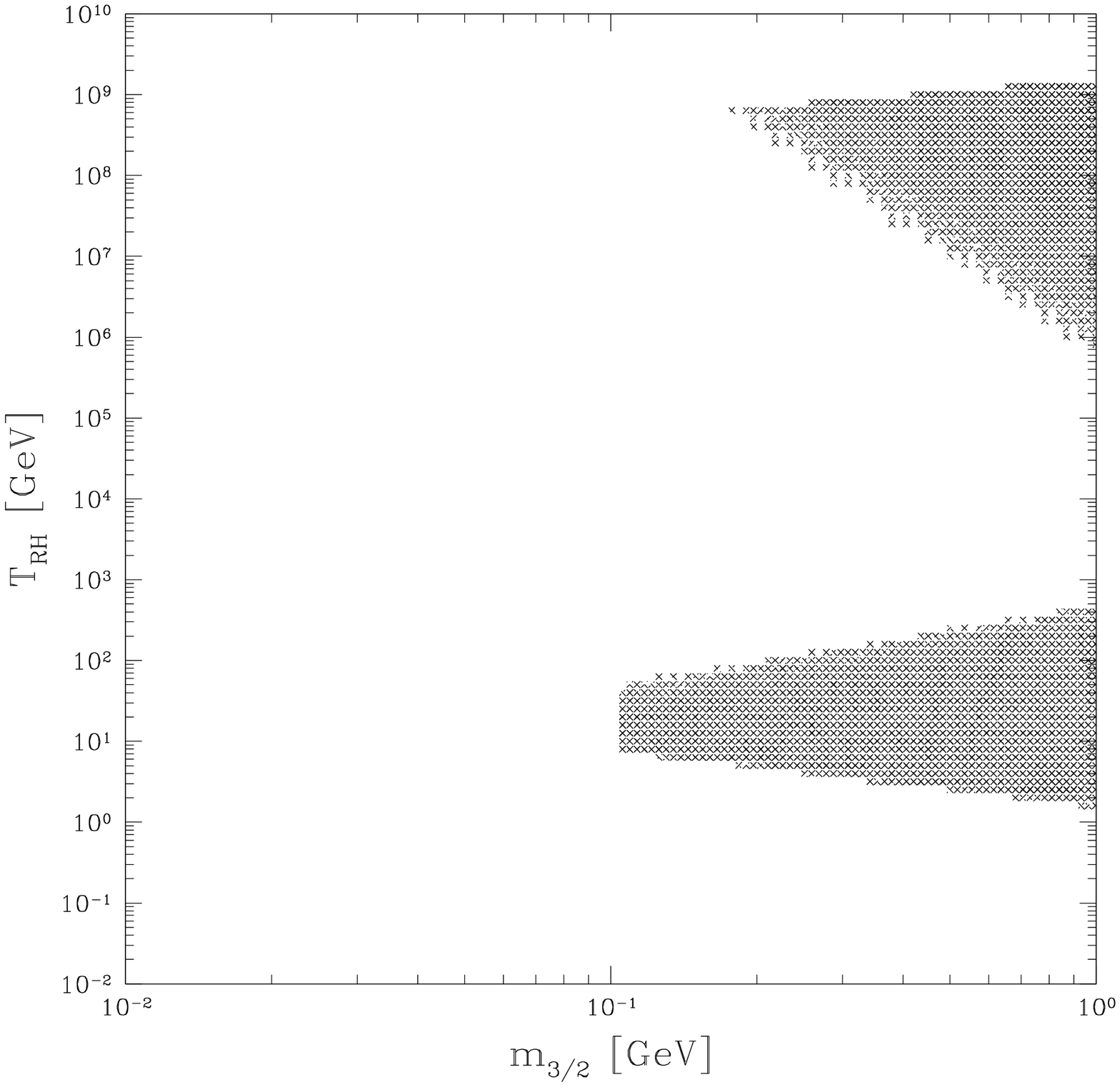} \includegraphics[width=10cm]{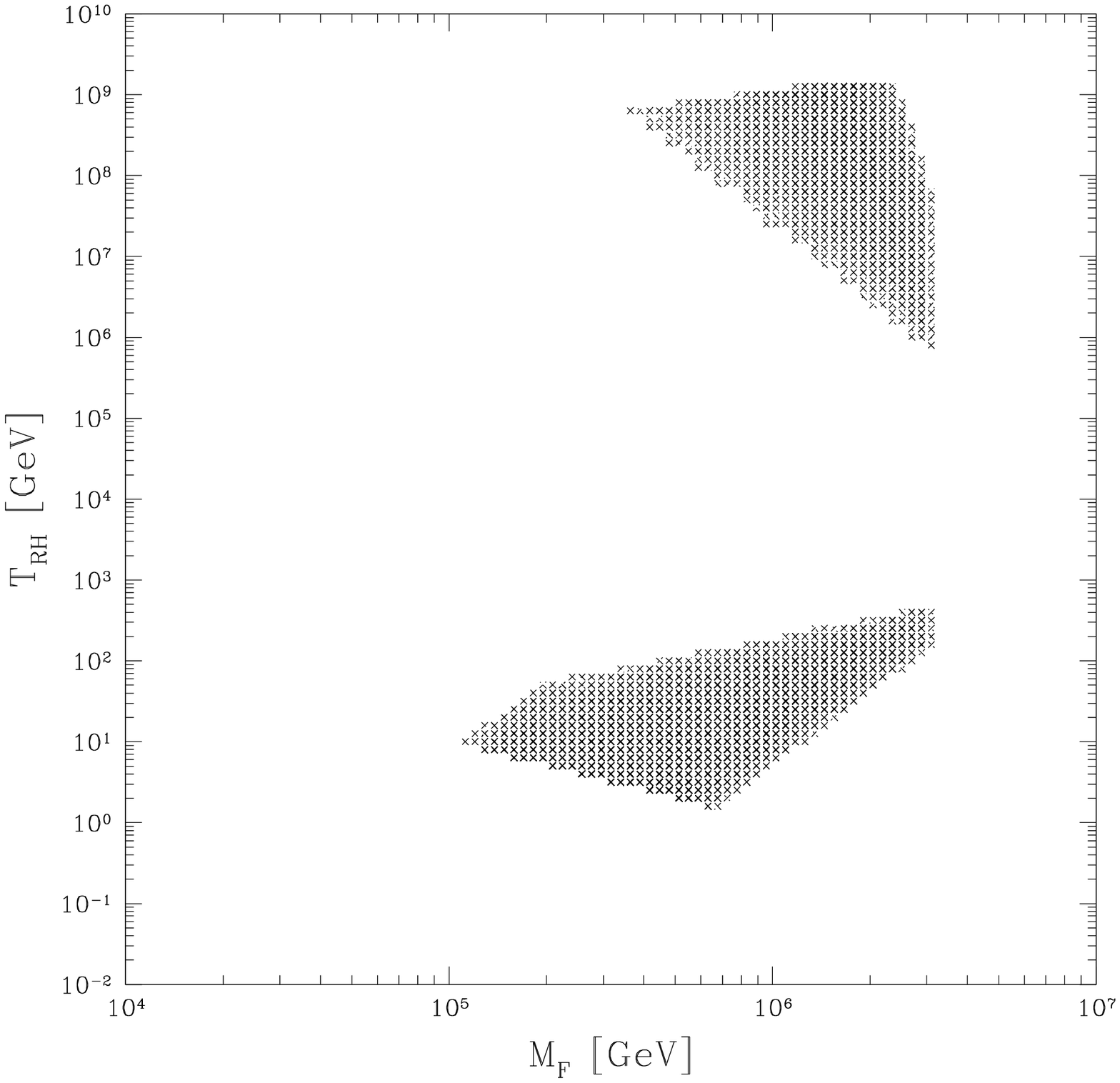}
\caption{\label{fig:region} The allowed region for $m_{3/2}$, $M_{F}$ and
$T_{RH}$, where our scenario succeeds and the baryon to entropy ratio
satisfies the following bounds : $10^{-11} \lesssim n_B/s \lesssim
10^{-10}$. 
Note that there does not exist any upper bound on $T_{RH}$ from the
gravitino problem~\cite{MMY, Gouvea}, since the L-balls dominate the universe
and their decay temperature is rather low.  The two separate allowed
region roughly corresponds to the cases A and B discussed in the text.  }
\end{figure}

The allowed regions for $m_{3/2}$, $M_{F}$, and $T_{RH}$ is shown
in Fig.~\ref{fig:region}, where the baryon to entropy ratio takes the
value required from  BBN,
\begin{equation}
\label{eq:nbscon}
10^{-11} \lesssim \frac{n_B}{s} \lesssim 10^{-10}.
\end{equation}
Here we adopt a rather loose constraint because of the uncertain CP
phase. As can be seen from Fig. \ref{fig:region}, there are two
allowed regions : (i) $m_{3/2} \sim 0.1 - 1$GeV, $M_{F}\sim 10^5 -
10^6$GeV, and $T_{RH}\sim 1 - 10^3$GeV, (ii) $m_{3/2} \sim 0.1 - 1$GeV,
$M_{F}\sim 10^6$GeV, and $T_{RH}\sim 10^6 - 10^9$GeV. Roughly speaking,
the regions (i) and (ii) correspond to the cases A and B respectively.

\begin{figure}
\includegraphics[width=13cm]{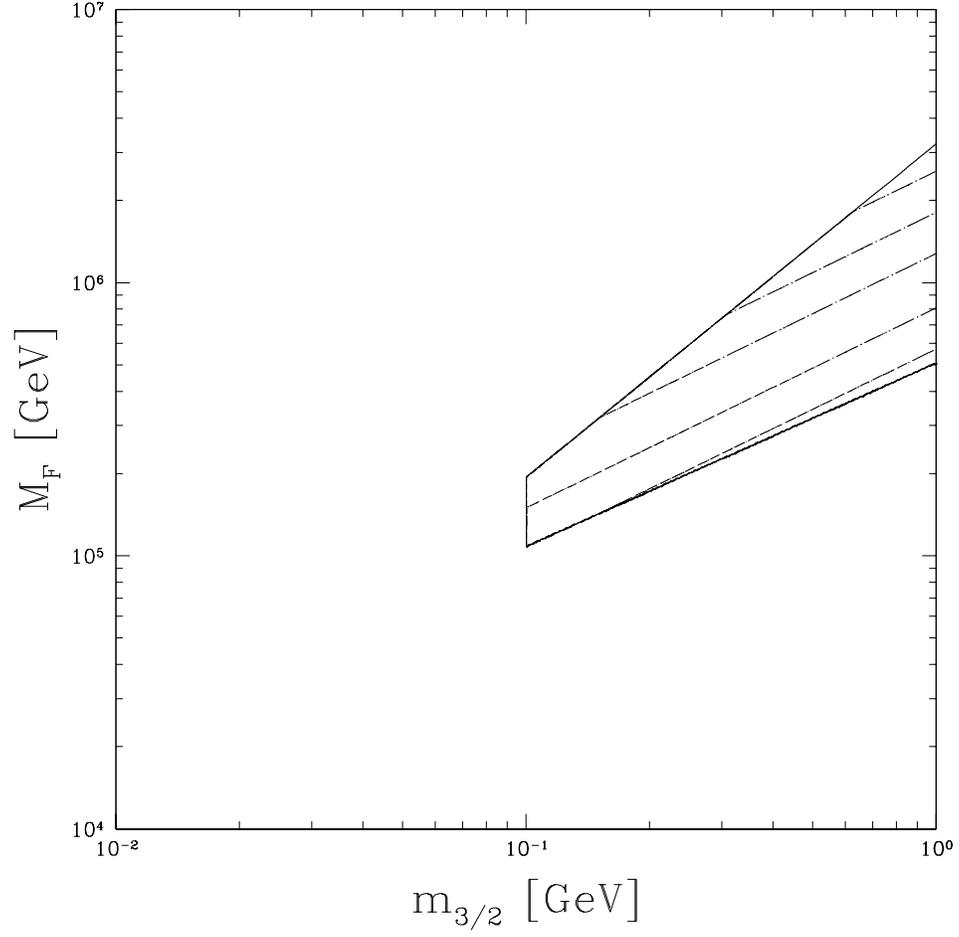} 
\caption{\label{fig:contour} 
The contours of the electron neutrino degeneracy are shown.  The
trapeziform area between two solid lines represents the allowed region
where our scenario works and the baryon to entropy ratio satisfies the
bounds: $10^{-11} \lesssim n_B/s \lesssim 10^{-10}$.  The contours
represents $\xi_{\nu_e} = 0.005,~0.01,~0.02,~0.06,~0.1$ from top to bottom.
}
\end{figure}

Also we plot the contours of the degeneracy of electron neutrinos in
Fig. \ref{fig:contour}, which shows that the large and positive lepton
asymmetry of electron type can be generated in our scenario. For
reference, the present constraint of $\xi_{\nu_e}$ by the analyses of
BBN and CMB data is given by \cite{Hansen},
\begin{eqnarray}
\label{eq:xi}
-0.01 \lesssim &\xi_{\nu_e}& \lesssim 0.22.
\end{eqnarray}
Thus, our scenario can generate both small baryon asymmetry and
positive large lepton asymmetry of electron type at the same time by
virtue of the AD leptogenesis and subsequently formed L-balls.

\section{Discussion and conclusions}

\label{sec:con}

In this paper we have proposed a scenario which accommodates small baryon
asymmetry and large lepton asymmetry simultaneously. The large lepton
asymmetry is generated through the Affleck-Dine mechanism and almost all
the produced lepton charges are absorbed into L-balls which are formed
subsequently. Thus, most of the produced lepton numbers do not suffer from
the sphaleron process. Only a small fraction evaporated from the L-balls due
to thermal effects is converted into baryon asymmetry, which is
responsible for the present baryon asymmetry.

As a concrete example, we consider positive large lepton asymmetry of
the electron type. The excess of electron neutrinos shifts the chemical
equilibrium between protons and neutrons toward protons so that the
predicted primordial abundance of $^{4}$He is decreased, which often gives
a solution to the discrepancy of BBN itself or that between BBN and
CMB. However, the sphaleron process converts lepton asymmetry into baryon
asymmetry with the opposite sign. To circumvent this problem, we identify
the $e^{c}_{1}L_{2}L_{3}$ flat direction to be the AD field. Then the
Affleck-Dine leptogenesis can generate positive lepton asymmetry of
the electron type but totally negative lepton asymmetry, which is converted
into positive baryon asymmetry. Of course, one should notice that by use
of another flat direction such as $e^{c}_{2}L_{1}L_{3}$, we can obtain
negative lepton asymmetry of the electron type and also total negative lepton
asymmetry.

Recently, it was pointed out that complete or partial equilibrium between
all active neutrinos may be accomplished through neutrino oscillations in
the presence of neutrino chemical potentials, depending on neutrino
oscillation parameters \cite{equilibrium}. In the case of partial equilibrium,
our scenario needs no change.  Only complete equilibrium can spoil our
scenario. Even if neutrino oscillation parameters lead to complete
equilibrium, our scenario may still work since it is possible that the
L-balls decay just before BBN and the complete equilibration cannot be
attained, which needs further investigation.

\subsection*{ACKNOWLEDGMENTS}

M.Y. was partially supported by the Japanese Grant-in-Aid for
Scientific Research from the Ministry of Education, Culture, Sports,
Science, and Technology.


\end{document}